\documentstyle[preprint,aps,epsf]{revtex}
\begin{document}
\draft

\title{Scaling and fractal formation in Persistence}

\author{ G.\ Manoj\cite{MAN} and P.\ Ray\cite{RAY}}

\address{ The Institute of Mathematical Sciences, C. I. T. Campus,
Taramani, Madras 600 113, India}

\date{\today}

\maketitle

\begin{abstract}
The spatial distribution of unvisited/persistent sites in $d=1$ $A+A\to\emptyset$
model is studied numerically. Over length scales smaller than
a cut-off $\xi(t)\sim t^{z}$, the set
of unvisited sites is found to be a fractal. The fractal dimension $d_{f}$,
dynamical exponent $z$ and persistence exponent $\theta$ are related
through $z(1-d_{f})=\theta$. The observed values of $d_{f}$ and $z$
are found to be sensitive to the initial density of particles. 
We argue that this may be due to the existence of two competing length
scales, and discuss the possibility of a crossover at late times.

\end{abstract}


\vspace{1cm}

Persistence properties of spatially extended systems undergoing 
time evolution has 
attracted a lot of attention of late. Generically there is a 
stochastic field
$\phi({\bf x},t)$ at each lattice site $\bf x$, which evolves
with time $t$ through interactions with other (usually nearest
neighbour) sites. One quantity of interest in the present context
is the persistence probability at time $t$, which is defined as the fraction
$P(t)$ of sites in which 
the stochastic field $\phi({\bf x},t)$ did not change sign in the
time interval [0,$t$]. In a large number of cases, it is found that
$P(t)\sim t^{-\theta}$ \cite{REVIEW}. The new exponent $\theta$, called 
persistence exponent, is, in general, unrelated to
other known static and dynamic exponents.

The non-trivial nature of $\theta$ can be
attributed to the interactions between neighbouring sites, which makes the
effective stochastic process at any single site non-Markovian.
Suppose $\phi({\bf x},t)$ flips sign at time $t$. This event
will increase the chance of neighbouring sites also flipping sign at
subsequent times $t^{\prime}> t$. This leads to the growth of 
spatial correlations in the system, which die out with increasing 
separation, on account of the statistical independence of distant flips.
The non-Markovian nature of the process is more directly captured in
these spatial correlations. Although much effort has been expended for
calculation of the persistence exponent $\theta$ by
exact\cite{DERRIDA} and approximate\cite{APPROX} methods, little has
been done to investigate the associated spatial
correlations in the process. In this paper, we undertake such a study
in $d=1$, where the correlations are expected to be most pronounced.

In one dimension, the zeroes of the stochastic field can 
be viewed as a set of particles, moving about in the lattice, annihilating each
other when two of them meet. When a particle moves across a lattice
site for the first time, the field there flips sign, and the
site becomes non-persistent. If each particle is assumed to perform
purely diffusive motion, this reduces to the well-known
reaction-diffusion model $A+A\to\emptyset$ \cite{PRIV}, 
with appropriate initial
conditions. The simplest case is random initial distribution of particles,
with average density $n_{0}$, for which $P(t)\sim t^{-\theta}$ 
with ${\theta}=3/8$\cite{DERRIDA}, independent of $n_{0}$
\cite{CARDY}. We investigate spatial correlations in persistence for this simple model.
We start with the two-point correlator $C(r,t)$, which is defined as
the probability that site {\bf x+r} is persistent, given that site
{\bf x} is persistent (averaged over {\bf x}).
We define $\rho({\bf x},t)$ as the density of persistent sites: 
ie., $\rho({\bf x},t)=1$ if site ${\bf x}$ is persistent at time $t$, and 0 otherwise. 
Then, with the previous definition, the expression for $C(r,t)$ is as follows.

\begin{equation}
C(r,t)= <\rho({\bf x},t)>^{-1}<\rho({\bf x},t)\rho({\bf x+r},t)>
\label{eq:CORR}
\end{equation}

where the brackets denote average over the entire lattice and
$<\rho({\bf x},t)>=P(t)$. 

Our main results are the following. Strong spatial
correlations exist in the distribution of persistent sites, with a 
cut-off length scale $\xi(t)$ separating correlated and
uncorrelated regions. At late times $t$ (ie., in the {\it scaling
regime}), this length scale grows as a 
power of time: $\xi(t)\sim t^{z}$, where $z$ is the dynamical
exponent in this context\cite{DYN}. In the correlated region $r\ll \xi(t)$, the
correlator shows a {\it power law decay with distance}:
$C(r,t)\sim r^{-\alpha}$. The scale-invariant behaviour, indicative of
strong correlations,
shows that the set of persistent sites is a self-similar fractal with dimension
$d_{f}=1-\alpha$. By consistency, the exponents are related as $z\alpha=\theta$.
We have analyzed the fractal structure by box-counting method also.
Careful measurements of the exponents over several
decades of MC time show that the observed values of $\alpha$ and $z$
change with initial density $n_{0}$ while satisfying the above scaling
relation. 

We did our numerical simulation
on a 1-d lattice of size $N= 10^{5}$, with
periodic boundary conditions.
Particles are initially distributed at random on the lattice with
average density $n_{0}$, and
their positions are sequentially updated--- each particle
was made to move one step in either direction with equal probability ($D=1/2$). 
Whenever such a move resulted in two particles occupying the
same position, both are removed from the lattice before moving to the
next particle. The starting density of persistent sites is
$P(0)=1-n_{0}$, and a persistent site becomes non-persistent when it is
occupied by a particle for the first time.
The time evolution is done up to $10^{5}$ Monte-Carlo steps (1 MC step
is counted after all the
particles in the lattice were touched once). These time and lattice
scales are the largest possible within our computational resources.
We repeated our simulations
for a few values of starting density $n_{0}$.
The results were averaged over 50 different initial realisations.

For distances $r\gg 1$ and late times $t$, we find that $C(r,t)\sim r^{-\alpha}$
for $r\ll \xi(t)$.
In the other extreme of large separations, ie., $r\gg \xi(t)$,
the sites are uncorrelated so that $C(r,t)=P(t)\sim t^{-\theta}$, independent of $r$. 
Thus $\xi(t)$ is the correlation
length for persistence, and consistency demands 
$\xi(t)^{-\alpha}\sim t^{-\theta}$. This implies a power-law
divergence: $\xi(t)\sim t^{z}$ with a
dynamical exponent $z$ related to $\alpha$ and $\theta$ through the
scaling relation

\begin{equation}
z\alpha=\theta
\label{eq:SCAL}
\end{equation} 

The observed behaviour of $C(r,t)$ can be summarised in the following
dynamic scaling form.

\begin{equation}
C(r,t)= P(t)f(r/\xi(t))
\label{eq:CORSCAL}
\end{equation}

with the scaling function $f(x)\sim x^{-\alpha}$ as 
$x\ll 1$ and $f(x)\simeq 1$ for $x\gg 1$. In Fig.1, the
scaling function $f(x)= C(r,t)/P(t)$ is plotted against the scaled
distance $x= r/t^{z}$ for two values of time separated by a decade.
The initial density is $n_{0}=0.5$. Excellent data collapse is
obtained for $z=1/2$, and
the measured value of the spatial exponent $\alpha\simeq 3/4$ is entirely
in accordance with the scaling relation.

The observed power-law decay of $C(r,t)$ with $r$ has a wider
significance, apart from showing the strong spatial correlations in
the distribution. It implies that, over length scales not too large, 
the underlying structure is a self-similar fractal. This is most easily
seen with the `box-counting' procedure \cite{STANLEY}.
We divide the enire lattice
into boxes of size $l$, at time $t$. After discarding `empty'
boxes, ie., those which contain not even a single persistent site, 
let $M(l,t)$ be the average number of persistent sites in a box of
length $l$. This quantity is related to $C(r,t)$ through 
$M(l,t)=\int_{0}^{l} C(r,t)dr$. Substituting the scaling form Eq.\
\ref{eq:CORSCAL} for
$C(r,t)$, one finds
 
\begin{equation}
M(l,t)\sim l^{1-\alpha} \hspace{0.5cm} l\ll \xi(t)
\end{equation}

\begin{equation}
M(l,t)=lP(t) \hspace{0.75cm} l\gg \xi(t)
\end{equation}

which can be summarised in the scaling form

\begin{equation}
M(l,t)= lP(t)h(l/\xi(t))
\label{eq:SCALMAS}
\end{equation}

with the scaling function $h(x)\sim x^{-\alpha}$ for $x\ll 1$ and
$h(x)\simeq 1$ for $x\gg 1$. We see that over small enough
length-scales $l\ll \xi(t)$, 
the set of persistent sites form a self-similar fractal with fractal dimension
$d_{f}=1-\alpha$, with a crossover to homogeneous behaviour at larger
length scales. This crossover is illustrated in Fig.2, where we have 
$M(l,t)$ (measured from box-counting) plotted against the box size $l$
for three values of time. The initial density here is $n_{0}=0.5$, and we find
$d_{f}\simeq 0.25$ in agreement with our result from study of the 
two-point correlation $C(r,t)$.

In Fig.3, we compare the results from box-counting for different starting
densities. For $n_{0}=0.2$, we see that the fractal region appears
much later compared to higher values. This is presumably due to the
large inter-particle separation at $t=0$, and the consequent delay in reaching
the scaling regime. For higher densities, the fractal dimension is
seen to decrease continously with $n_{0}$, approaching zero in the limit
$n_{0}\to 1$. We notice that although $\xi(t)\sim 10^{3}$ in terms of the
lattice spacing, it is still much less than the lattice size $N$,
so as to rule out finite-size effects.

In Fig.4, we plot
the scaling function $h(\eta)=M(l,t)/lP(t)$ 
against the scaling variable $\eta =l/t^{z}$ for two values of time
separated by a decade. We have displayed results for $n_{0}=0.8$ and
0.95. For $n_{0}=0.8$ the best data collapse
is obtained with $z\simeq 0.45$, wheras for $n_{0}=0.95$, the
corresponding value is $z\simeq 0.39$. The exponent $\alpha$, measured
from the small argument divergence of $h(\eta)$, also shows similar
changes.

In Table I, we have summarised our exponent values for four initial densities. 
All measurements were made using the data for the mass-distribution $M(l,t)$
rather than the correlator $C(r,t)$ on account of lesser statistical fluctuations.
For the dynamical exponent $z$, we chose the value which gave the best
collapse of data under dynamic scaling. Although it is difficult to measure
the exponent very accurately using this method, we have verified by visual
inspection that the error involved is less than the reported
changes in the exponent values at least by a factor of two. We have
omitted the case $n_{0}=0.2$ because no single value of $z$ was found
to give good scaling behaviour in the time range studied. 

A more direct way to measure the dynamical exponent $z$ is to compute
the average separation $L(t)$ between persistent sites. If the spatial distribution
were uniform over all length scales, this quantity would be simply $L(t)\sim
P(t)^{-1}$. Since this is not the case, we have to proceed more carefully.  
We define $n(k,t)$ to be the number
of nearest neigbour pairs of persistent sites at time $t$ with
separation $k$. 
By definition, $\int_{k}n(k,t)= NP(t)$ and $\int_{k}kn(k,t)=N$.
The average separation $L(t)=N^{-1}\int_{k}k^{2}n(k,t)$ and we expect
$L(t)\sim t^{z}$. 
We computed $L(t)$ numerically by simulating
100 lattices of size $N=10^{5}$ upto $10^{5}$ time steps, for each $n_{0}$.
In Fig.5, we display the results for the running exponent
$z_{eff}=d($log$ L)/d($log$ t)$. The results are seen to be fully supportive 
of our earlier conclusions.

Our numerical results are strongly suggestive of non-universal
behaviour of exponents $\alpha$ and $z$. The non-universal
exponent values have been observed to be valid over at least three decades of
MC time (upto $10^{5}$ time steps). We note that
there are {\it two length scales} at work
here. For low $n_{0}$, the dynamics is dominated by
diffusive motion of isolated
particles, `eating into' clusters of persistent sites. Due to annihilation,
their average density decays as $n(t)=(8\pi Dt)^{-1/2}$ \cite{LUSH} and hence
the average separation is the diffusive scale ${\cal
L}_{D}(t)\sim t^{1/2}$. 
On the other hand, for $n_{0}\to1$, the initial separation of persistent sites 
$\sim 1/(1-n_{0})\gg 1$. The short time behaviour is
now dominated by persistent $\to$ non-persistent conversion of isolated 
sites, with characteristic length scale ${\cal L}_{p}(t)\sim t^{3/8}$.
It is possible that the observed non-universal behaviour
results from competition between these two scales. According to this
picture, one should see a crossover to diffusion dominated regime at
later times, but we are
yet to see any signature of that. Further numerical work, at least a
few orders of magnitude greater than what is reported here, would be required to
establish conclusively the possibility of a temporal crossover.

To conclude, we have discovered strong, power-law correlations in the
spatial distribution
of persistent sites in one-dimensional $A+A\to\emptyset$ model. The correlation length
$\xi(t)$ exhibits an algebraic divergence with time. In the correlated
region, the set of persistent sites form a
self-similar fractal, while over larger length scales, the distribution is
homogeneous. These features are not specific to this model or dimension.
We have observed identical features in kinetic Ising model in $d=1$
and 2 \cite{FUTURE}, showing that this is a general phenomenon in the context of
persistence. The interesting
aspect of the present model is that the fractal dimension was found to
be sensitive to the starting density of particles. 

We thank M. Muthukumar for discussions and G. I. Menon for a critical
reading of the manuscript and suggestions.


\begin{table}
\begin{tabular}{ccc}
$n_{0}$ & $\alpha$ & $z$ \\
\hline
0.50 & 0.7342(8) & 0.50 \\
0.80 & 0.8294(5) & 0.45 \\
0.95 & 0.9517(3) & 0.39 \\
\end{tabular}
\narrowtext
\vspace{0.5cm}
\caption { Observed values of exponent $\alpha$ as measured from
box-counting method (details in text), for four values of initial density
$n_{0}$. The quoted value of dynamical exponent $z$ is the one which
gave the best data collapse over three decades of time, 
$t=10^{3}, 10^{4}$ and $10^{5}$.  
The fractal dimension $d_{f}=1-\alpha$.
}
\label{tab:TAB1}
\end{table}

\newpage
\begin{center}
FIGURE CAPTIONS
\end{center}

FIG. I: The scaling function for two-point correlation $f(x)=C(r,t)/P(t)$
plotted against the scaling variable $x=r/t^{z}$ on log-scale for two values
of time $t=10^{3}$ and $10^{4}$. The starting density of particles
is $n_{0}=0.5$. The data for different times are seen to collapse into
the same curve if scaling is done with $z=0.50$. The observed 
$\alpha \simeq 0.75$ is in agreement with the
proposed scaling relation (2). 
\vspace{1.0cm}

FIG. II: The average number of persistent sites $M(l,t)$ in a box of size $l$
at time $t$ is plotted against the box size $l$ for
$t= 10^{3}, 10^{4}$ and $10^{5}$. The initial density of
particles is $n_{0}=0.5$. The crossover from fractal (dimension
$d_{f}\simeq 1/4$) to homogeneous ($d_{f}=d=1$) distribution is clear
from the figure.

\vspace{1.0cm}

FIG. III: Same as Fig.2, for four starting densities $n_{0}=0.2, 0.5, 0.8$ and 0.95.
All plots correspond to $t=10^{4}$. For $n_{0}=0.2$, the fractal
region is reached late, but the asymptotic value is seen to be the
same as that for $n_{0}=0.5$. For higher $n_{0}$, $d_{f}$ decreases
continously, approaching zero in the limit $n_{0}\to 1$.
\vspace{1.0cm}

FIG. IV: The scaling function for the mass-distribution $h(\eta)=M(l,t)/lP(t)$
plotted against the scaling variable $\eta=r/t^{z}$ for two values
of time $t=10^{4},10^{5}$ and two starting densities $n_{0}=0.8$ and
0.95. The observed data collapse has been obtained with
$z=0.45 (n_{0}=0.8)$ and $z=0.39 (n_{0}=0.95)$. The corresponding
values for $\alpha$ are $\simeq$ 0.83 and 0.95. For comparison, a
straight line with slope 0.75 is also shown.

\vspace{1.0cm}
FIG. V: The running exponents $z_{eff}$ for four starting densities 
is plotted against $1/$log$ t$. These results have been averaged over
100 starting configurations.


\begin{references}
\bibitem{DYN} We have chosen a different convention for defining $z$
from what is generally found in literature.
\end{references}
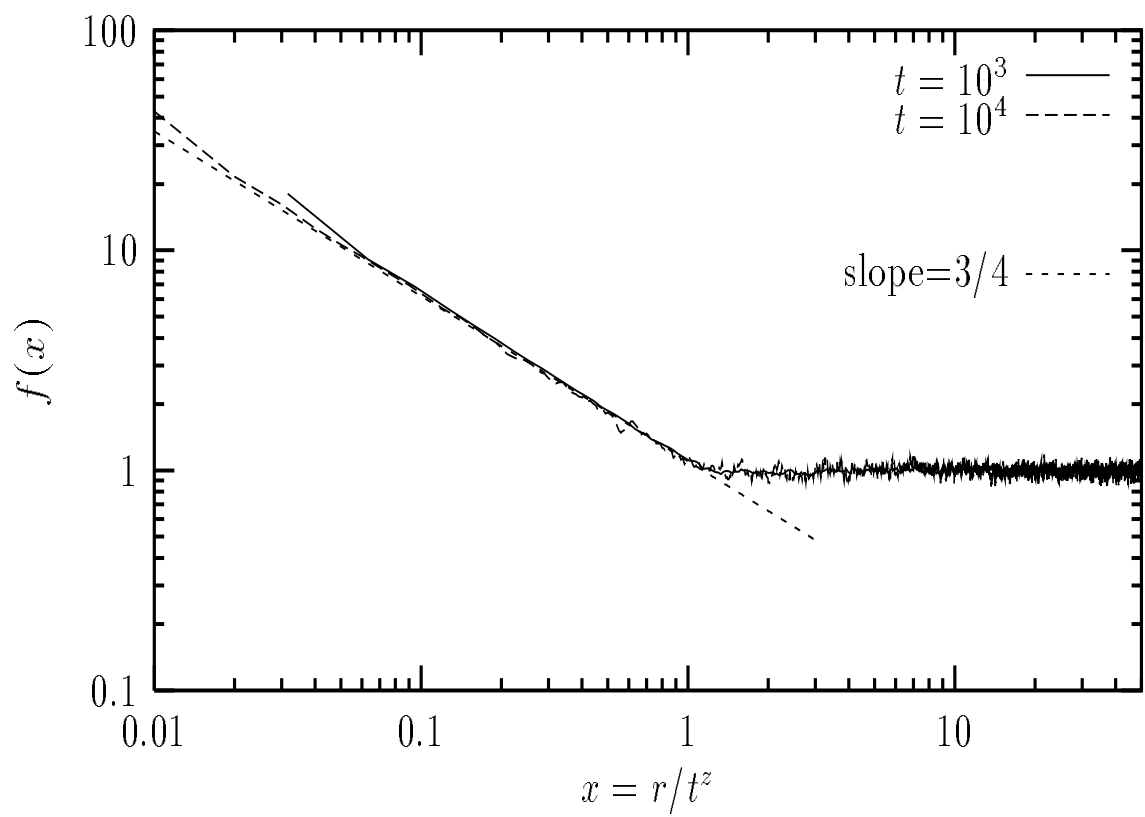
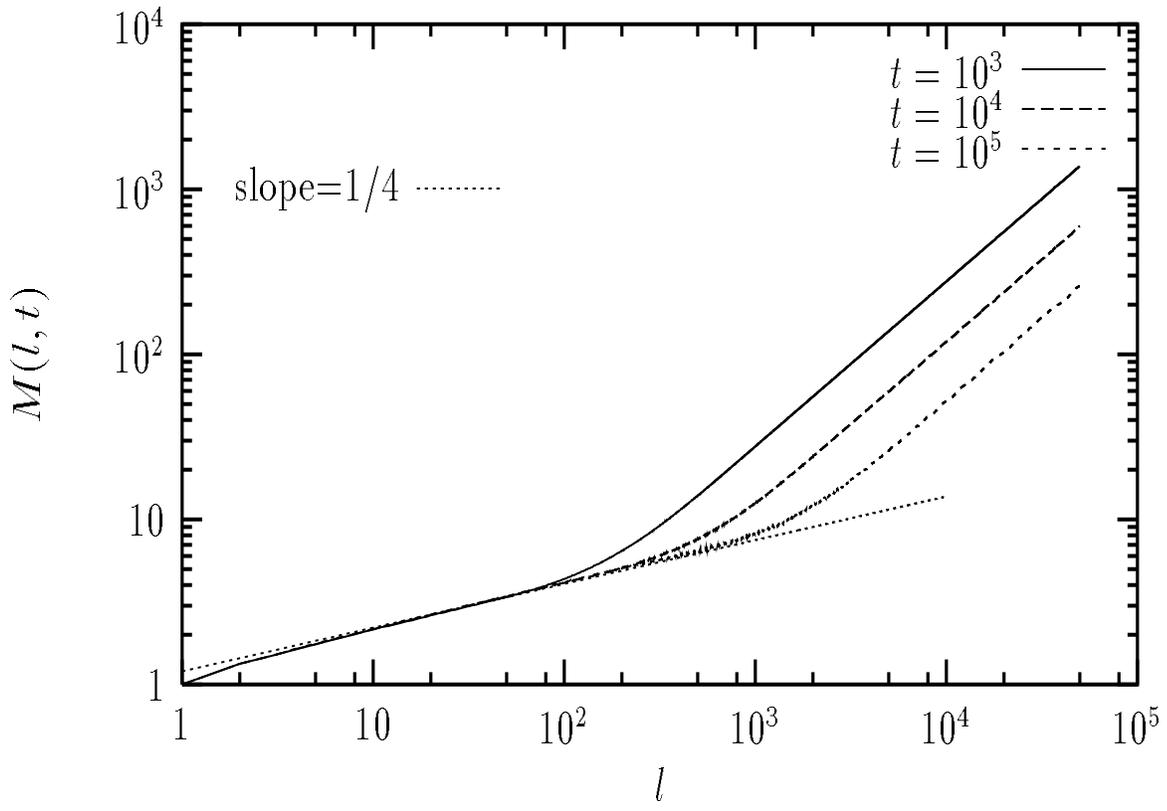
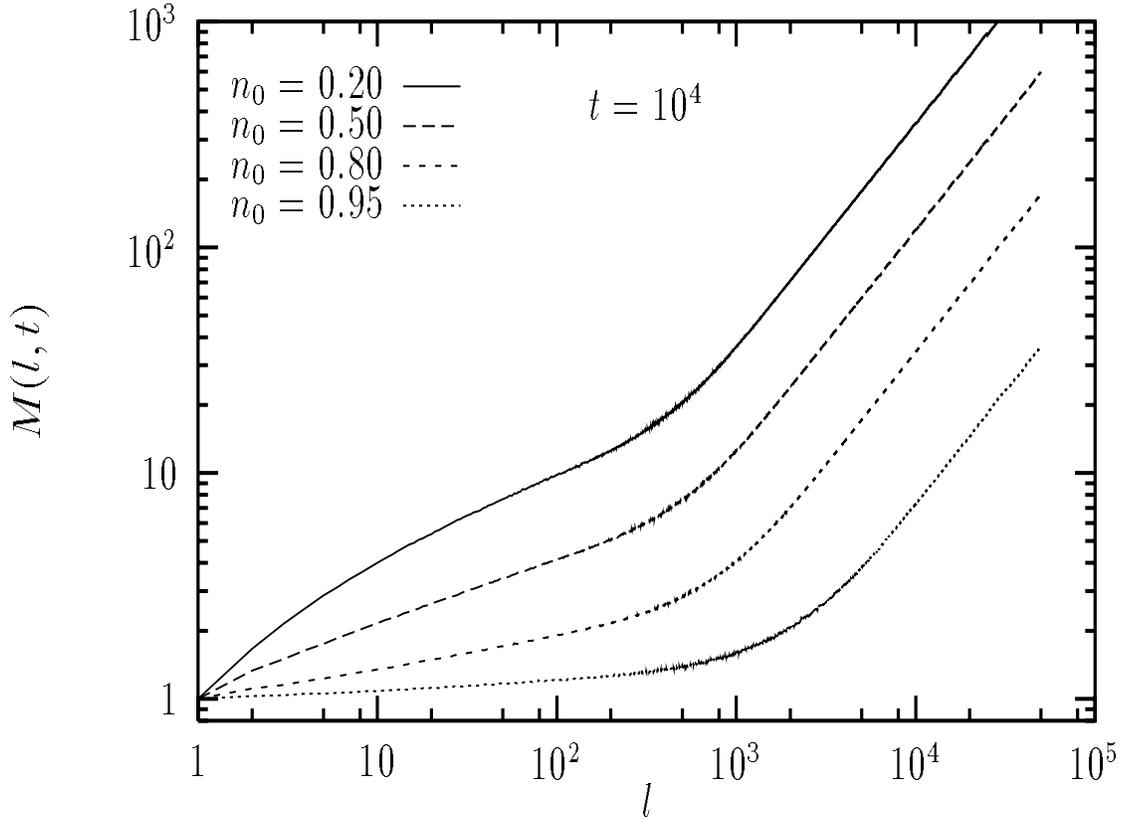
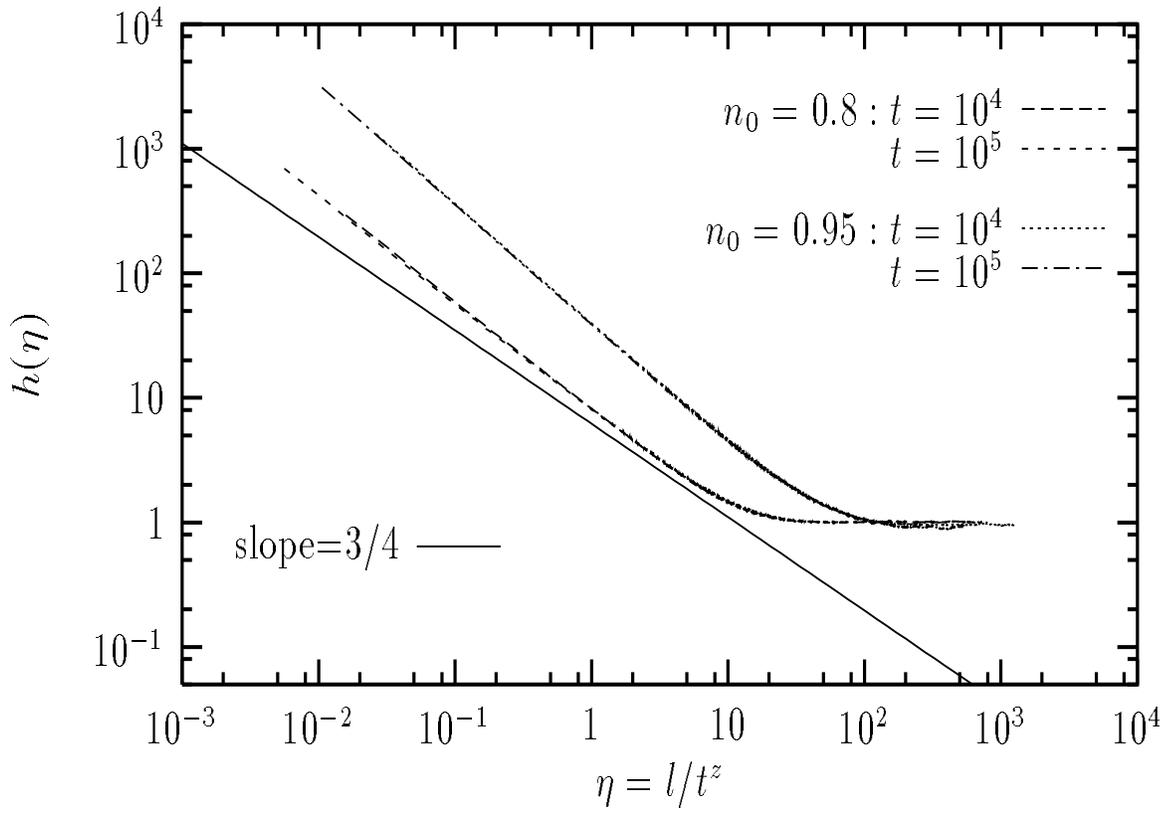
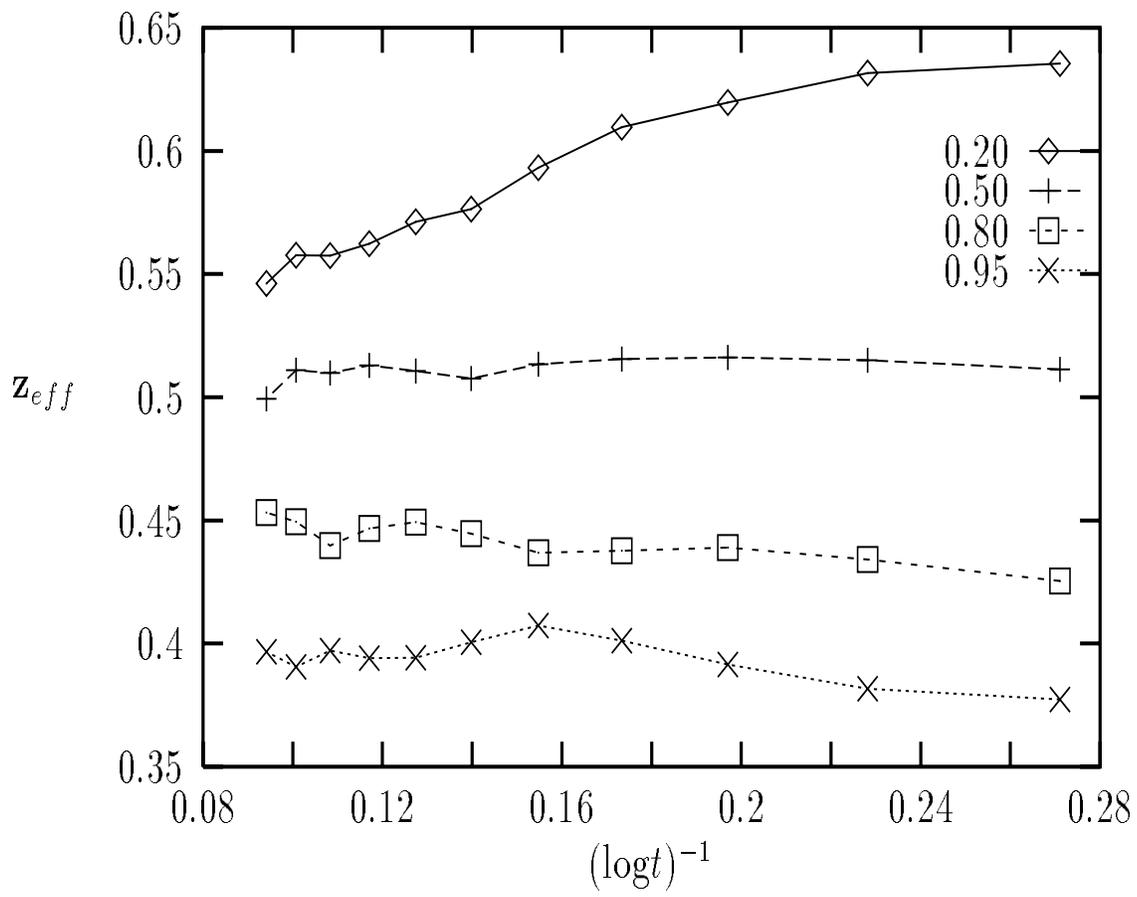
\end{document}